\documentclass[12pt]{article} \input epsf.tex

\newcommand{\be}{\begin{equation}}
\newcommand{\ee}{\end{equation}}
\newcommand{\bear}{\begin{eqnarray}}
\newcommand{\eear}{\end{eqnarray}} \newcommand{\ba}{\begin{array}}
\newcommand{\ea}{\end{array}}
\newcommand{\lae}{\begin{array}{c}\,\sim\vspace{-21pt}\\<
\end{array}}
\newcommand{\gae}{\begin{array}{c}\,\sim\vspace{-21pt}\\>
\end{array}}

\newcommand{\CA}{{\cal A}} \newcommand{\CQ}{{\cal Q}}
\newcommand{\CU}{{\cal U}} \newcommand{\CD}{{\cal D}}
\newcommand{\CL}{{\cal L}} \newcommand{\CE}{{\cal E}}

\begin{document}

\pagestyle{empty} \begin{titlepage}
\def\thepage {}        

\title{\LARGE \bf Bounds on Universal Extra Dimensions \\ [1cm]}

\author{
\small\bf \hspace*{-.3cm} Thomas Appelquist$^1$, Hsin-Chia
Cheng$^2$, Bogdan A.~Dobrescu$^{1}$ \\ \\ {\small {\it
\hspace*{-.6cm} $^1$Department of Physics, Yale University, New
Haven, CT 06511, USA\thanks{e-mail: thomas.appelquist@yale.edu,
hcheng@theory.uchicago.edu, bogdan.dobrescu@yale.edu}}}\\
 {\small {\it $^2$Enrico Fermi
Institute, The University of Chicago, Chicago, IL 60637, USA }}\\
 }

\date{ } \maketitle

   \vspace*{-8.9cm}
\noindent \makebox[11.8cm][l]{\small \hspace*{-.2cm}
hep-ph/0012100} {\small YCTP-P12-00 } \\
\makebox[11.8cm][l]{\small \hspace*{-.2cm} December 8, 2000 }
{\small EFI--2000-48} \\

 \vspace*{9.2cm}

  \begin{abstract}
{\small We show that the bound from the electroweak data on the
size of extra dimensions accessible to all the standard model
fields is rather loose. These ``universal'' extra dimensions could
have a compactification scale as low as 300 GeV for one extra
dimension. This is because the Kaluza-Klein number is conserved
and thus the contributions to the electroweak observables arise
only from loops. The main constraint comes from weak-isospin
violation effects. We also compute the contributions to the $S$
parameter and the $Zb\bar{b}$ vertex.
The direct bound on the compactification scale is set by CDF and
D0 in the few hundred GeV range, and the Run II of the Tevatron
will either discover extra dimensions or else it could
significantly raise the bound on the compactification scale. In
the case of two universal extra dimensions, the current lower
bound on the compactification scale depends logarithmically on the
ultra-violet cutoff of the higher dimensional theory, but can be
estimated to lie between 400 and 800 GeV. With three or more extra
dimensions, the cutoff dependence may be too strong to allow an
estimate. }

\end{abstract}

\vfill \end{titlepage}

\baselineskip=18pt \pagestyle{plain} \setcounter{page}{1}


\section{Introduction} \setcounter{equation}{0}

Extra dimensions accessible to standard model fields are of
interest for various reasons. They could allow gauge coupling
unification \cite{dudas}, and provide new mechanisms for
supersymmetry breaking \cite{Antoniadis:1990ew} and the generation
of fermion mass hierarchies \cite{Arkani-Hamed:2000dc}. More
recently it has been shown that extra dimensions accessible to the
observed fields may lead to the existence of a Higgs doublet
\cite{Arkani-Hamed:2000hv}.

A number of studies indicate that if standard model fields
propagate in extra dimensions, then they must be compactified at a
scale $1/R$ above a few TeV \cite{bounds}. These studies refer,
however, to theories in which some of the quarks and leptons are
confined to flat four-dimensional slices (branes). In the
equivalent four-dimensional theory where the extra dimensions are
accounted for by towers of heavy Kaluza-Klein (KK) states, the
bound on $1/R$ is due to the tree level contributions of the KK
modes to the electroweak observables.

In this paper we point out that extra dimensions accessible to
{\it all\/} the standard model fields, referred to here as
universal dimensions, may be significantly larger. The key element
is the conservation of momentum in the universal dimensions. In
the equivalent four-dimensional theory this implies KK number
conservation. In particular there are no vertices involving only
one non-zero KK mode, and consequently there are no tree-level
contributions to the electroweak observables. Furthermore,
non-zero KK modes may be produced at colliders only in groups of
two or more. Thus, none of the known bounds on extra dimensions
from single KK production at colliders or from electroweak
constraints applies for universal extra dimensions.

The heavy KK modes contribute, though, at loop-level to the
electroweak observables, so that some lower bound on $1/R$ can be
set. In addition, there is a direct bound on $1/R$ from the
non-observation of KK pair production at the Tevatron and LEP.
After presenting some general features of universal extra
dimensions in section 2, we compute the bound on their size from
the electroweak data (section 3.) We then discuss the current
direct bound on $1/R$ from collider experiments (section 4.) Our
conclusions are summarized in section 5.

\section{The Kaluza-Klein spectrum and interactions}
\setcounter{equation}{0}

Our starting point is the minimal standard model in $d = 4 +
\delta$ space-time dimensions. The gauge, Yukawa and quartic-Higgs
couplings have negative mass dimension, so this is an effective
theory, valid below some scale $M_s$. We assume a compactification
scale $1/R < M_s$ for the $\delta$ extra spatial dimensions. The
upper bound on $1/R$ for the class of models being discussed is
determined by the range of validity of the effective 4-dimensional
Higgs theory. To avoid fine-tuning the parameters in the Higgs
sector, $1/R$ should not be much higher than the electroweak
scale. We study the experimental lower bound on $1/R$. Given that
the gauge couplings and the top Yukawa coupling are of order one
at the electroweak scale, the $d$-dimensional theory remains
perturbative for a range of energies above $1/R$. The cutoff $M_s$
on the $d$-dimensional theory is expected to be no higher than the
upper end of this range.

We use the generic notation $x^\alpha, \; \alpha = 0, 1, ..., 3
+\delta$ for the coordinates of the $(4+\delta)$-dimensional
space-time, but we explicitly distinguish between the usual
non-compact space-time coordinates, $x^\mu, \, \mu = 0,1,2,3$, and
the coordinates of the extra dimensions, $y^a, \, a = 1, ...,
\delta$. The 4-dimensional Lagrangian can be obtained by
dimensional
reduction from the $(4+\delta)$-dimensional theory, 
\bear {\cal L}(x^\mu) & = & \int d^{\delta} y \left\{ -
\sum_{i=1}^3 \frac{1}{2 \hat{g}_i^2} {\rm
Tr}\left[F_i^{\alpha\beta}(x^\mu, y^a)
{F_i}_{\,\alpha\beta}(x^\mu, y^a)\right]  + {\cal L}_{\rm
Higgs}(x^\mu,y^a) \right. \nonumber \\ [2mm] &+& \left. \!\!\!\! i
\left(\overline{\CQ}, \overline{\CU}, \overline{\CD}\right)
(x^\mu, y^a) \left(\Gamma^\mu D_\mu + \Gamma^{3+a} D_{3+a} \right)
\left(\CQ, \CU, \CD \right)^\top (x^\mu, y^a) \right.
\nonumber \\ [2mm] &+& \left. \!\!\!\!\!
\left[\overline{\CQ}(x^\mu,y^a)\left(\hat{\lambda}_\CU
\CU(x^\mu,y^a) i\sigma_2 H^*(x^\mu,y^a) +
 \hat{\lambda}_\CD \CD(x^\mu, y^a) H(x^\mu, y^a)\right) +
{\rm h.c.} \right] \right\} ~. \label{lagrangian} \eear  
Here $F^{\alpha\beta}_i$ are the $(4+\delta)$-dimensional gauge
field strengths associated with the $SU(3)_C \times SU(2)_W \times
U(1)_Y$ group, while $D_\mu = \partial/\partial x^\mu - \CA_\mu$
and $D_{3+a} = \partial/\partial y^a - \CA_{3+a}$ are the
covariant derivatives, with $\CA_\alpha = -i \sum_{i=1}^3\hat{g}_i
{\CA_\alpha^{r}}_i T^{r}_i$ being the $(4+\delta)$-dimensional
gauge fields. The piece ${\cal L}_{\rm Higgs}$ of the
$(4+\delta)$-dimensional Lagrangian contains the kinetic term for
the $(4+\delta)$-dimensional Higgs doublet $H$, and the Higgs
potential. The $(4+\delta)$-dimensional gauge couplings
$\hat{g}_i$, and the
 Yukawa couplings collected in the $3\times 3$ matrices
 $\hat{\lambda}_{\CU,\CD}$, have dimension (mass)$^{-\delta/2}$.

The fields $\CQ, \CU$ and $\CD$ describe the
$(4+\delta)$-dimensional fermions whose zero-modes are given by
the 4-dimensional standard model quarks. A summation over a
generational index is implicit in Eq.~(\ref{lagrangian}). For
example, the 4-dimensional, third generation quarks may be written
as $\CQ^{(0)}_3 \equiv (t, b)_L, \, \CU^{(0)}_3 \equiv t_R, \,
\CD^{(0)}_3 \equiv b_R$. The kinetic and Yukawa terms for the
weak-doublet and -singlet leptons, $\CL$ and $\CE$, are not shown
for brevity.

The gamma matrices in $(4+\delta)$ dimensions, $\Gamma^\alpha$,
are anti-commuting $2^{k + 2}\times 2^{k + 2}$ matrices, where $k$
is an integer such that $\delta = 2k$ or $\delta = 2k+1$. Chiral
fermions exist only when $\delta$ is even, and correspond to the
eigenvalues $\pm 1$ of $\Gamma^{4+\delta}$.
Therefore, if the spacetime has an odd number of dimensions
($\delta = 2k+1$), $\CQ, \CU, \CD, \CL,$ and $\CE$ are vector-like
$2^{k + 2}$-component fermions, and the $(4+\delta)$-dimensional
theory is automatically anomaly-free.  For an even number of
dimensions ($\delta =2k$) one may choose $\CQ, \CU, \CD, \CL$ and
$\CE$ to be chiral $2^{k + 1}$-component fermions. In order to
have Yukawa couplings with the scalar Higgs field, the
$SU(2)_W$-doublet fermions and the $SU(2)_W$-singlet fermions must
have opposite chiralities. This guarantees that the unbroken
$SU(3)_C$ and $U(1)_{\rm EM}$ are vector-like, hence anomaly free.
The gravitational anomaly may easily be cancelled by gauge-singlet
fermions. The $SU(2)_W$ and $U(1)_Y$ gauge groups are chiral, so
there can be $(4+2k)$-dimensional anomalies involving the
$SU(2)_W$ and $U(1)_Y$ gauge groups, but they can be cancelled by
the Green-Schwarz mechanism~\cite{Green:1984sg}. For both odd and
even $\delta$ the 4-dimensional anomalies cancel because the
fermion content is chosen so that the effective theory at scales
below $1/R$ is the 4-dimensional standard model.

In order to derive the 4-dimensional Lagrangian from
Eq.~(\ref{lagrangian}), we must specify the compactification of
the extra dimensions. The simplest choice is an $[(S^1 \times
S^1)/Z_2]^k$ orbifold for $\delta = 2k$, and an $[(S^1 \times
S^1)/Z_2]^k\times (S^1/Z_2)$ orbifold for $\delta = 2k+1$. An
orbifold of this type is a $\delta$-dimensional torus cut in half
along each of the $y^a$ coordinates with odd $a$.
 Each component of a $d$-dimensional field that belongs to a
 representation of the 4-dimensional Lorentz group, $SO(3,1)$,
 must be either odd or even under
 the orbifold projection: $(y^a, \, y^{a+1}) \rightarrow (-y^a,\,
-y^{a+1})$ for even $a + 1 \le 2k $, as well as $y^{2k+1}
\rightarrow -y^{2k+1}$ for $\delta = 2k +1$. An equivalent
description of the compactification is a $\delta$-dimensional
space with coordinates $0 \le y^a \le \pi R$ for {\it odd\/} $a$
and $-\pi R \le y^a \le \pi R$ for {\it even\/} $a$, and boundary
conditions such that each field or its derivatives with respect to
the $y^a$'s vanish at the orbifold fixed points $y^a=0,\, \pm \pi
R$. ($\Phi =0,\,\partial^2 \Phi/\partial y^a \partial y^b=0$ for
odd fields, and $\partial \Phi/\partial y^a =0$ for even fields at
the orbifold fixed points.)

The Lagrangian (\ref{lagrangian}) together with the boundary
conditions completely specifies the theory.
For $\delta = 2$, the $SU(3)_C \times SU(2)_W \times U(1)_Y$ gauge
fields are decomposed in KK modes as follows: 
\bear \CA_\nu
(x^\mu,y^a) & = & \frac{\sqrt{2}}{(2\pi R)^{\delta/2}} \left\{
\CA_\nu^{(0,0)}(x^\mu)_{ \ba{c} \\ \ea } \hspace*{-1em} \right. +
\sqrt{2} \sum_{j_1,j_2}
 \left.
\CA_\nu^{(j_1,j_2)}(x^\mu)_{\ba{c} \\ \ea} \!\!\!\! \cos\left[
\frac{1}{R}(j_1 y^1 + j_2 y^2) \right] \right\} ~,
 \nonumber \\ [2mm]
\CA_b (x^\mu,y^a) & = & \frac{2}{(2\pi R)^{\delta/2}} 
\sum_{j_1,j_2}
\CA_b^{(j_1,j_2)}(x^\mu) \, \sin\left[ \frac{1}{R}(j_1 y^1 + j_2
y^2) \right] \label{bosons} ~,
\eear
where the summation is over all integer values of the KK numbers
$j_1$ and $j_2$ 
that satisfy $j_1 + j_2 \ge 1$, or $j_1 = -j_2 \ge 1$.
The gauge fields polarized in the $x^\nu, \, \nu = 0, 1,2,3$
directions are even under the orbifold transformation, so that the
zero-modes correspond to the 4-dimensional standard model gauge
fields. On the other hand, the gauge fields polarized along the
coordinates $y^b, \, b= 1,2$ of the extra dimensions are odd under
the orbifold transformation, so that their zero-modes are
projected out and no massless scalar fields appear after
dimensional reduction.

With boundary conditions in the $\delta =2$ compact dimensions
chosen to give the appropriate chiral structure for the KK
zero-modes, the KK decomposition for the top quark fields is given
by \bear \CQ_3 (x^\mu,y^a) & = &  \frac{\sqrt{2}}{(2\pi
R)^{\delta/2}}\left\{ (t,b)_L(x^\mu)_{ \ba{c} \\ \ea }
\hspace{-1em} \right. 
+ \sqrt{2} \sum_{j_1, j_2} \left[ P_L
{\CQ_3}_L^{(j_1,j_2)}(x^\mu) \, \cos\left( \frac{1}{R}(j_1 y^1 +
j_2 y^2) \right) \right. \nonumber \\ [0.5em] && + \left.\left.
P_R{\CQ_3}_R^{(j_1,j_2)}(x^\mu) \, \sin\left( \frac{1}{R}(j_1 y^1
+ j_2 y^2) \right)\right] \right\} ~,
 \nonumber \\ [1em]
\CU_3 (x^\mu,y^a) & = & \frac{\sqrt{2}}{(2\pi R)^{\delta/2}}
\left\{ t_R(x^\mu)_{ \ba{c} \\ \ea } \hspace{-1em} \right. 
+ \sqrt{2} \sum_{j_1, j_2} \left[ P_R
{\CU_3}_R^{(j_1,j_2)}(x^\mu) \, \cos\left( \frac{1}{R}(j_1 y^1 +
j_2 y^2) \right) \right. \nonumber \\ [0.5em] && + \left.\left.
P_L{\CU_3}_L^{(j_1,j_2)}(x^\mu) \, \sin\left( \frac{1}{R}(j_1 y^1
+ j_2 y^2) \right) \right] \right\} \label{quarks} ~, \eear 
where the range of values for  $j_1$ and $j_2$ 
is the same as above, in Eq.~(\ref{bosons}).
The third-generation weak-doublet quark, $\CQ_3 = (\CQ_t, \CQ_b)$, 
and weak-singlet up-type quark, $\CU_3 $, are four-component 
chiral fermions in six dimensions so that their KK modes are
4-dimensional vector-like quarks, with the exception of the
zero-modes which are chiral. The chiral projection operators that
appear in the KK decomposition, $P_{L,R} = (1 \mp \gamma_5)/2$,
are the 4-dimensional ones.

For $\delta =1$,  the KK decomposition may be obtained from
Eqs.~(\ref{bosons}) and (\ref{quarks}) by setting $j_2 = 0, \, y^2
= 0$. In general, for $\delta = 2k+1$ one may compactify one
dimension as above and then compactify the remaining $k$ pairs of
extra dimensions, with appropriate use of the higher dimensional
chiral projection operators. For $\delta = 2k$, the KK
decomposition may also be obtained by iterating
Eqs.~(\ref{bosons}) and (\ref{quarks}) $k$ times. The other quark
and lepton $d$-dimensional fields have similar KK-mode
decompositions. The Higgs field must be even under the orbifold
transformation. Only the Higgs zero-mode acquires an electroweak
asymmetric vacuum expectation value (VEV) of $v/\sqrt{2} \approx
174$ GeV.

The heavy spectrum in four dimensions consists of KK levels
characterized by the mass eigenvalues \be M_j = \frac{p_j}{R}
~,\ee where $j\ge 1$, $p_{j+1} > p_j$, and $p_j$ is given by \be
j_1^2 + ... + j_\delta^2 = p_j^2 ~. \ee The degeneracy of the
$j$th KK level, $D_j$, is given by the number of solutions to this
equation for $j_1, ..., j_\delta$. At each level there would be
$D_j$ sets of fields, each of them including the $SU(3)_C \times
SU(2)_W \times U(1)_Y$ gauge fields, three generations of {\it
vector-like\/} quarks and leptons, a Higgs doublet, and $\delta$
scalars in the adjoint representations of $SU(3)_C \times SU(2)_W
\times U(1)_Y$.

 An essential observation is that the momentum
conservation in the extra dimensions, implicitly associated with
the Lagrangian (\ref{lagrangian}), is preserved (as a discrete
symmetry) by the above orbifold projection. In the case of
fermions, this implies that there is no mixing among the modes of
different KK levels. The zero-mode top-quark gets a mass from its
Yukawa coupling exactly as in the 4-dimensional standard model.
Given that each KK level includes a tower of both left- and
right-handed modes for each of the $t_L$ and $t_R$ fields, there
is a $2\times 2$ mass matrix for each top-quark KK level. The
$D_j$ top-quark mass matrices of the $j$th KK level may be written
in the weak eigenstate basis as \be \left( \overline{\CU}_3^j ,
\overline{\CQ}_t^j \right) \left(\ba{c c} -M_j & m_t \\ m_t & M_j
\ea \right) \left( \ba{c} \CU^j_3 \\ \CQ_t^j \ea \right) ~.
\label{matrix}
 \ee
Here, $\CQ_t^j$ is a four-component field describing the
 $j$th KK modes associated with $t_L$. The diagonal
terms are the masses induced by the kinetic terms in the $y^a$
directions, while the off-diagonal terms are the contributions
from the Higgs VEV. The corresponding mass eigenstates,
$\CU_3^{\prime j}$ and $\CQ_t^{\prime j}$, have the same mass, \be
M^{(j)}_t = \sqrt{M_j^2 + m_t^2} ~. \ee The weak eigenstate top KK
modes are related to the mass eigenstates by
 \be
\left( \ba{c} \CU^j_3 \\ \CQ_t^j \ea \right) = \left(\ba{c c} -
\gamma_5 \cos\alpha_j  & \sin\alpha_j  \\ \gamma_5 \sin\alpha_j  &
\cos\alpha_j  \ea \right) \left( \ba{c} \CU^{\prime j}_3 \\
\CQ^{\prime j}_t \ea \right) ~, \label{matrix2}
 \ee
where $\alpha_j$ is the mixing angle,
 \be \tan 2\alpha_j =
\frac{m_t}{M_j}~. \label{alpha} \ee

The $b$-quark KK modes have an analogous structure due to the
mixing between $\CD_3^j$ (the left- and right-handed fields
associated with $b_R$) and $\CQ_b^j$ (the left- and right-handed
fields associated with $b_L$). However, this mixing may be
neglected up to corrections of order $(m_b/m_t)^2$ compared with
the mixing from the top-quark KK sector.

The interactions of the KK modes may be derived from the
$d$-dimensional Lagrangian (\ref{lagrangian}), using the KK
decompositions shown in Eqs.~(\ref{bosons}) and (\ref{quarks}),
and by integrating over the extra dimensions. For the one-loop
computations to be considered here, it is sufficient to know the
vertices involving one or two zero-modes and two non-zero modes.
In the remainder of this section we list the relevant terms of
this type that appear in the 4-dimensional Lagrangian.

The top and bottom mass-eigenstate KK modes ({\it i.e.}, the
vector-like quarks $\CQ_t^{\prime j}, \, \CU_3^{\prime j}$,
$\CQ_b^j$ and $\CD_3^j$) have the following electroweak
interactions with the $W^\pm$ and $Z$ zero-modes:
 \bear {\cal
L}_{W_1} & = & \frac{g}{2 \cos \theta_W}Z_\mu \left[ \left(
\sin^2\alpha_j - \frac{4}{3} \sin^2\theta_W \right)
\overline{\CU}_3^{\prime j} \gamma^\mu \CU_3^{\prime j} + \left(
\cos^2\alpha_j - \frac{4}{3} \sin^2\theta_W \right)
\overline{\CQ}_t^{\prime j} \gamma^\mu \CQ_t^{\prime j} \right.
 \nonumber \\ [2mm] && \left.
+ \sin\alpha_j \cos\alpha_j \left(\overline{\CU}_3^{\prime j}
\gamma^\mu \gamma_5 \CQ_t^{\prime j} + {\rm h.c.} \right)
+ 2 g_L^b \overline{\CQ}_b^j \gamma^\mu \CQ_b^j + 2 g_R^b
\overline{\CD}_3^j \gamma^\mu \CD_3^j \right] \nonumber \\ [2mm] &
+ & \frac{g}{\sqrt{2}} \left[ W^+_\mu \left( - \sin\alpha_j
\overline{\CU}_3^{\prime j} \gamma_5 + \cos \alpha_j
\overline{\CQ}_t^{\prime j} \right) \gamma^\mu \CQ_b^j + {\rm
h.c.} \right] ~,
\label{interaction} \eear where \be g_L^b = - \frac{1}{2} +
\frac{1}{3} \sin^2\theta_W \; , \; \; g_R^b = \frac{1}{3}
\sin^2\theta_W  ~. \label{coupling} \ee

 The weak eigenstate neutral gauge bosons,
${W^{3}_\mu}^j$ and $B^{j}_\mu$ mix level by level in the same way
as the neutral $SU(2)_W$ and hypercharge gauge bosons in the
4-dimensional standard model. The corresponding mass eigenstates,
$Z_\mu^j$ and $A^j_\mu$, have masses $\sqrt{M_j^2 + M_Z^2}$ and
$M_j^2$, respectively. These heavy gauge bosons have interactions
with one zero-mode quark and one $j$-mode quark (in the
weak-eigenstate basis) identical with the standard model
interactions of the zero-modes.

Likewise, there are interactions of one quark zero-mode and one
quark $j$-mode with the $j$-mode of the scalars corresponding to
the electroweak gauge bosons polarized along $y^a$, $W_{a+3}^j,\,
Z_{a+3}^j, \, A_{a+3}^j$. For $\delta = 1$, they may be
written as follows: 
\bear {\cal L}_{W_2} \hspace*{-.5em} & = &
\hspace*{-.5em}
 \frac{e}{3} i A_4^j \left[ 2\cos\alpha_j \left(
\overline{\CQ}_{t_R}^{\prime j}t_L - \overline{\CU}_{3_L}^{\prime
j} t_R \right) - 2\sin\alpha_j \left(\overline{\CU}_{3_R}^{\prime
j} t_L  - \overline{\CQ}_{t_L}^{\prime j}t_R \right) -
\overline{\CQ}_{b_R}^j b_L - \overline{\CD}_{3_L}^{j} b_R \right]
\nonumber \\ [2mm] & + & \hspace*{-.8em}
 \frac{g}{\cos\theta_W} i Z_4^j
\left[ \overline{\CQ}_t^{\prime j} \left( c^j_{1_V} + c_{1_A}^j
\gamma_5 \right) t + \overline{\CU}_3^{\prime j} \left( c_{2_V}^j
+ c_{2_A}^j \gamma_5 \right) t + g_L^b\CQ_{b_R}^j b_L + g_R^b
\overline{\CD}_{3_L}^{j} b_R \right] \nonumber \\ [2mm] &+&
\hspace*{-.8em} \frac{g}{\sqrt{2}} i {W^+_4 }^j \left(
\cos\alpha_j\overline{\CQ}_{t_R}^{\prime j} b_L - \sin\alpha_j
\overline{\CU}_{3_R}^{\prime j} b_L + \overline{t}_L \CQ_{b_R}^j
\right) + {\rm h.c.} 
\eear 
where 
\be 
c_{1_{V,A}}^j = \pm
\cos\alpha_j \left(\frac{1}{4} - \frac{1}{3} \sin^2\theta_W
\right) - \frac{1}{3} \sin\alpha_j \sin^2\theta_W ~, \label{cav}
\ee 
and $c_{2_{V,A}}^j$ are obtained by permuting $\sin\alpha_j$
and $\cos\alpha_j$ in the above expression. For $\delta = 2$ the
$W_{a+4}^j,\, Z_{a+4}^j, \, \gamma_{a+4}^j$ scalars have similar
couplings, up to sign differences, while for $\delta \ge 3$ the
gauge bosons polarized along each pair of extra dimensions couples
to a different set of quark KK modes.

Each non-zero KK mode of the Higgs doublet, $H^j$, includes a
charged Higgs and a neutral CP-odd scalar of mass $M_j$, and also
a neutral CP-even scalar of mass $\sqrt{M_j^2+M_h^2}$. The
interactions of the Higgs and gauge boson KK modes may also be
obtained from the corresponding standard model interactions of the
zero-modes by replacing two of the fields at each vertex with
their $j$th KK mode.

\section{Electroweak data versus extra dimensions}
\setcounter{equation}{0}

We study the sensitivity of the electroweak observables to the
higher dimensional physics setting in at scale $1/R$. The largest
contributions come from the KK modes associated with the
top-quark, but there are also corrections due to the gauge and
Higgs KK modes. QCD corrections are small and are neglected. The
standard model in universal extra dimensions is described by four
unknown parameters: the Higgs boson mass $M_h$, the
compactification radius $R$, the cutoff scale $M_s$, and the
number of extra dimensions $\delta$. The upper bound on the cutoff
$M_s$ is determined in terms of $1/R$ and the value of the various
couplings at this scale. The Higgs boson mass is bounded from
above by the requirement that the Higgs quartic coupling,
$\lambda_h$, does not blow up (from a perturbative point of view)
at a scale significantly below $M_s$. We use this constraint ($M_h
\lae 250$ GeV \cite{Arkani-Hamed:2000hv}), and study the lower
bound on $1/R$, concentrating on the observables that are most
likely to yield severe constraints.

An important question is whether the electroweak observables can
be computed within the framework of the effective, higher
dimensional theory, that is, whether they are sensitive to the
unknown physics at scale $M_s$ and above. We will show that in the
case of one extra dimension we can reliably ignore the effects of
KK modes heavier than the cut-off (see section 3.1).  With two
extra dimensions the KK modes give corrections to the electroweak
observables that depend logarithmically on the cut-off, and in
more extra dimensions the dependence is more sensitive (see
section 3.2).

Given that the large $t-b$ mass splitting requires a hierarchy
between the top and bottom Yukawa couplings which in turn induces
weak isospin violation in the KK spectrum, the parameter \be
\Delta\rho \equiv\alpha T = \Delta\left( \frac{M_W}{M_Z
\cos\theta_W} \right), \ee which measures the splitting in the $W$
and $Z$ masses due to physics beyond the standard model, is a
prime suspect for constraining $1/R$.

The one-loop contribution to $\Delta \rho$ from one KK level
associated with the $t$ and $b$ quarks follows from
Eq.~(\ref{interaction}) and is given by \be \alpha T_j^t =
\frac{3m_t^2}{16 \pi^2 v^2}
 f_T\left(m_t^2/M_j^2\right) ~,
\ee where $v = 246$ GeV and
 \bear f_T(z) & = & 1 - \frac{2}{z} + \frac{2}{z^2} \ln \left( 1 + z
\right) \nonumber \\ [2mm] & = & \frac{2z}{3} \left[ 1
-\frac{3z}{4} + \frac{3z^2}{5}
 + {\cal O} \left(z^3\right) \right]
~.\eear The form of this contribution to $\Delta\rho$ is easy to
understand. The factor $m_t^2 /(4 \pi v)^2$ arises from the
definition of $\Delta\rho$ as the coefficient of the
lowest-dimension, weak isospin-violating term in the electroweak
chiral Lagrangian \cite{Appelquist:1993ka}. The additional factor
$( m_t^2 / M_j^2)$ is present because the non-zero KK modes
decouple in the large mass limit.

In addition to the top and bottom KK modes, the Higgs KK modes
contribute to $\Delta\rho$ because the VEV of the zero-mode Higgs
induces isospin violation in the couplings of the higher modes of
the Higgs doublet. To leading order in $M_h^2/M_j^2$, one Higgs KK
mode gives \be \alpha T^h_j = -
\left(\frac{\alpha}{4\pi\cos^2\theta_W}\right) \frac{5 M_h^2 + 7
M_W^2}{12 M_j^2} \ee  Finally, the KK electroweak gauge bosons
also contribute, giving \be \alpha T_j^V = -
\left(\frac{\alpha}{4\pi\cos^2\theta_W}\right) \frac{(2\delta +
11) M_W^2 }{6 M_j^2} ~. \ee In each of these expressions, the
factor $\alpha /\cos^2\theta_W$ is present because the hypercharge
gauge interaction provides the weak-isospin symmetry breaking. The
second factor is present because the higher KK modes decouple in
the large mass limit.

The contribution to $T$ from all the KK modes is
 \be
T = \sum_{j=1}^{n_{\rm max}} D_j \left(T_j^t + T_j^h + T_j^V
\right)~. \label{sum} \ee The upper limit $n_{\rm max}$
corresponds to the mass scale $M_s$ at which the effective
$d$-dimensional theory breaks down. Note that the total number of
contributing KK modes of a particular field is \be N_{\rm KK} =
\sum_{j=1}^{n_{\rm max}} D_j ~. \ee Using the experimental values
for $M_W$, $M_Z$, $m_t$, and $\alpha$, the $T$ parameter may be
written in the form \bear T \approx  0.76 \sum_{j=1}^{n_{\rm max}}
D_j \frac{m_t^2}{M_j^2} && \!\!\!\!\! \left\{1 - 0.81
\frac{m_t^2}{M_j^2} + 0.65 \frac{m_t^4}{M_j^4} + {\cal
O}\left(m_t^6/M_j^6\right) \right. \nonumber \\ [2mm] && \left. -
0.057 \frac{M_h^2}{m_t^2} \left[1 + {\cal
O}\left(M_h^2/M_j^2\right) \right] \right\}  ~,
\label{T-parameter} \eear where $m_t \approx 175$ GeV. The error
here is about 10\% due to uncertainties in $m_t$, higher terms in
$M_h$, and the range of values of $\delta$ being considered. The
current upper bound on $T$ is approximately 0.4 at 95\% CL for
$M_h \lae 250$ GeV (and is somewhat relaxed for larger $M_h$
\cite{Chivukula:2000px}.) The experimental bound on $1/R$ is a
function of the KK spectrum, which depends on the number of extra
dimensions. We return to this estimate in section 3.1.

In addition to the $T$ parameter, the corrections from new physics
to the electroweak gauge boson propagators are encoded in the $S$
parameter defined by: \be S \equiv - \frac{8 \pi}{M_Z^2} \left(
\Pi_{3Y}(M_Z^2) - \Pi_{3Y}(0) \right) \label{expv} \\  ~,
\label{Sdef} \ee where $\Pi_{3Y}(q^2)$ is the vacuum polarization
induced by non-standard physics (note that the gauge couplings are
factored out according to the definition for hypercharge where $Y
\equiv 2(Q - T_3)$). The $S$ parameter gets a one-loop
contribution from each top-quark KK level: \be S_j^t =  -
\frac{1}{2 \pi} \int_0^1 d x \left\{ 3 \frac{m_t^2}{M_Z^2} \ln
\left[ 1 - \frac{x(1-x)M_Z^2}{M_j^2 + m_t^2} \right] + 2 x(1-x)
\ln \left[ 1 + \frac{m_t^2}{M_j^2 - x(1-x)M_Z^2}\right] \right\}
~. \ee Assuming that
 $M_j^2 \gg m_t^2$, this expression takes the form \be
S_j^t \approx \frac{1}{12 \pi} \frac{m_t^2}{M_j^2} \left[ 1 -
\frac{m_t^2}{M_j^2} \left(2 + \frac{ M_Z^2 }{10m_t^2 }\right) +
\frac{m_t^4}{M_j^4} \left(\frac{7}{3} - \frac{ M_Z^2}{5 m_t^2} +
\frac{3 M_Z^4}{70m_t^4} \right) + {\cal O}
\left(m_t^6/M_j^6\right) \right] ~.\ee

The Higgs KK mode contribution to $S$ is given, to leading order
in $M_h^2/M_j^2$, by \be S_j^h = \frac{ M_h^2 + \left( -3
-2\cos\theta_W^2\right)M_Z^2}{24 \pi M_j^2} ~, \ee while the gauge
boson KK modes do not contribute. Using the experimental values
for $M_Z$, $m_t$, and $\cos \theta_W$, the total contribution to
$S$ may be written in the form
 \bear
S  = \sum_{j=1}^{n_{\rm max}} D_j \left(S_j^t + S_j^h \right) &&
\nonumber \\ [2mm]  \!\!\!\!\approx 10^{-2} \sum_{j=1}^{n_{\rm
max}} D_j \frac{m_t^2}{M_j^2} \;\; && \!\!\!\!\!\!\!\! \left\{ 1 -
5.4 \frac{m_t^2}{M_j^2} + 6.0 \frac{m_t^4}{M_j^4} + {\cal
O}(m_t^6/M_j^6)\right. \nonumber \\ [2mm] && \left.
 + 1.3\frac{M_h^2}{m_t^2}\left[1 + {\cal O}(M_h^2/M_j^2) \right]
\right\} ~, \label{sumS} \eear where $m_t \approx 175$ GeV. This
result is smaller by almost two orders of magnitude than the one
for $T$, assuming that the series in $m_t^2/M_j^2$ and
$M_h^2/M_j^2$ are convergent. Given that the bounds on $S$ and $T$
are comparable ($S \lae 0.2$ at 95\% CL), we see that once the
bound on $1/R$ from $T$ is satisfied there is no relevant
constraint from $S$. This is not surprising because the quark KK
modes are vector-like fermions and therefore contribute to $S$
only if their masses violate the custodial symmetry, which leads
to a large $T$.

Another potential constraint on $1/R$ arises due to the one-loop
corrections of the KK modes to the $Z \rightarrow b\bar{b}$
branching ratio. The vertex correction is usually encoded in the
quantity \be \Delta R_b = 2 R_b(1 - R_b) \frac{ g_{L}^b \Delta
g_{L}^b + g_{R}^b \Delta g_{R}^b} {(g_{L}^b)^2 + (g_{R}^b)^2} \ee
where $R_b$ is the ratio of the $Z$ decay widths into $b \bar{b}$
and hadrons, while $g_{L,R}^b$ appear in the standard model $Z b
\bar{b}$ couplings at tree-level, and are given in
Eq.~(\ref{coupling}).

The contribution to $\Delta g_{L}^b$ due to top-quark KK modes,
for $M_j^2 \gg m_t^2$, is given by: \bear \Delta g_L^b & = &
\frac{\alpha}{ 4\pi}\left(\frac{1}{2\sin^2\theta_W} -1 \right)
\sum_{j=1}^{n_{\rm max}} D_j \frac{m_t^2}{M_j^2} \nonumber \\
[2mm] &\approx & 7.2 \times 10^{-4} \sum_{j=1}^{n_{\rm max}} D_j
\frac{m_t^2}{M_j^2} ~. \eear There are also corrections from Higgs
and gauge boson KK modes, but they are significantly smaller. The
contribution to $\Delta g_R^b$ is suppressed by $m_b^2/M_j^2$ and
may also be neglected. The standard model prediction is $R_b^{\rm
SM} = 0.2158$, so that $\Delta R_b \approx - 0.77 \Delta g_{L}^b$,
while the measured value is $R_b^{\rm exp} = 0.21653 \pm 0.00069$
\cite{ewwg}. Notice that the correction to $R_b$ from the KK modes
has the wrong sign, and therefore is tightly constrained. The
$2\sigma$ bound is $\Delta R_b > - 7 \times 10^{-4}$, which gives
$\Delta g_{L}^b < 9.4 \times 10^{-4}$. One can then derive a bound
for $\sum D_j/M_j^2$, but it is easy to see (for $M_j^2 \gg
m_t^2$) that this is less severe than the bound imposed by the $T$
parameter.

The shift in $g_{L}^b$ also affects the left-right asymmetry
measured by SLD, which depends on \be A_b \equiv \frac{(g_{L}^b)^2
- (g_{R}^b)^2} {(g_{L}^b)^2 + (g_{R}^b)^2} ~. \ee The correction
due to the KK modes is given by $\Delta A_b \approx - 0.29 \Delta
g_{L}^b$. Using the SM prediction, $A_b^{\rm SM} = 0.935$, and the
measured value $A_b^{\rm exp} = 0.922 \pm 0.023$ \cite{ewwg}, one
can easily check that this constraint is much looser than the one
from $R_b$.

We  expect that all other electroweak observables impose no
stronger constraints on $1/R$ than the one from the $T$ parameter.

\subsection{Bounds on one universal extra dimension}

We consider first the case of a single extra dimension. Then $D_j
= 1$ and $M_j = j/R$, with $R$ the compactification radius, so
that the summations over KK modes in Eq.~(\ref{T-parameter}) are
convergent. Extending the sums to $n_{\rm max} \gg 1$ gives
 \be T \approx 1.2 (m_t R)^2 \left[ 1 - 0.53(m_t R)^2 + 0.40
(m_t R)^4+ {\cal O} \left(m_t^6 R^6\right) \right] ~. \ee The
current upper bound on isospin breaking effects, $T \lae 0.4$,
yields a lower bound on the compactification scale: \be
\frac{1}{R} \gae 300 \; {\rm GeV} ~. \ee

The $S$ parameter and other electroweak observables  also involve
convergent KK mode sums in the 5-dimensional case. As discussed
above, they are less constraining than the $T$ parameter.

The convergence of each of these mode sums indicates that the
electroweak observables can indeed be computed reliably within the
effective 5-dimensional theory, relevant below the cutoff $M_s$.
The convergence of the computations can be understood by recalling
that each is effectively a 5-dimensional integral -- a
4-dimensional integral plus a KK mode sum. The convergence of the
corresponding 4-dimensional integrals for the electroweak
observables is well known, and this is not changed with only a
single additional dimension.

The reliability of these computations and the consequent lower
bound $1/R \gae 300 \; {\rm GeV}$ can be checked by examining
higher order corrections in the effective 5-dimensional theory. In
the limit $M_s \gg 1/R$, the 5-dimensional couplings become strong
at the cutoff, and there are potentially large corrections to the
one-loop result. Consider the two loop corrections, for example.
The integrals are now logarithmically divergent, but there are two
additional powers of a 5-dimensional coupling, each of which is
proportional to $1/\sqrt{M_s}$. Thus these corrections have a
suppression factor of $1/(RM_s)$ relative to the one-loop
estimate. Higher loops can all be seen to be of this order,
meaning that within the effective 5-dimensional theory the
corrections to the one-loop results can only be estimated.
Nevertheless, they are all suppressed by the factor $1/(RM_s)$,
indicating the same for the unknown physics above $M_s$. When
$M_s$ is well below the scale where the 5-dimensional couplings
become strong, the higher loops may be ignored. The unknown
physics above the cutoff induces effective operators in the
$d$-dimensional theory suppressed by powers of $M_s$. After
dimensional reduction the corresponding 4-dimensional operators
are further suppressed by powers of $1/(RM_s)$.

To estimate the largest value of $M_s$ below which the theory is
perturbative, we note that the loop expansion parameters can be
written in the form \be \epsilon_i = N_i
\frac{\alpha_i(M_s)}{4\pi} N_{\rm KK}(M_s)~, \label{epsilon} \ee
where the $\alpha_i$ are the 4-dimensional standard model gauge
couplings, the index $i=1,2,3$ labels the $U(1)_Y$, $SU(2)_W$ and
$SU(3)_C$ groups, $N_i = 1,2,3$ is the corresponding number of
colors, and $N_{\rm KK}(M_s)$ is the number of KK modes below
$M_s$. The value of $M_s$ at which these parameters become of
order unity is the largest cutoff consistent with a perturbative
effective theory.
 Each of the $D_j$ sets of fields within one KK
level contributes to the one-loop coefficients of the three
$\beta$ functions an amount $(81/10, \, 4/3 , \, -5/2)$. Although
the 4-dimensional $SU(3)_C$ coupling becomes more asymptotically
free above each KK level, the $d$-dimensional $SU(3)_C$
interaction becomes non-perturbative in the ultraviolet before the
other gauge interactions. The $\epsilon_3$ parameter becomes of
order unity, indicating breakdown of the effective theory, at
roughly $10$ TeV. The KK modes above that scale, as well as
operators induced by other physics above the cutoff, give
negligible contributions to the electroweak observables.

\subsection{Two or more universal extra dimensions}

For $d \geq 6$, the $T$ and $S$ parameters, and other electroweak
observables become cutoff dependent. The KK mode sums diverge in
the limit $N_{\rm KK} \rightarrow \infty$ because the KK spectrum
is denser than the 5-dimensional case.  This can again be seen by
noting that the 4-dimensional integrals plus the KK mode sums are
effectively $d$-dimensional integrals. The electroweak observables
($S, T, ...$), convergent in four and five dimensions at one loop,
become logarithmically divergent at $d = 6$ and more divergent in
higher dimensions. The degeneracies $D_j$ and masses $M_j$ of the
KK modes are listed for a toroidal compactification in
Ref.~\cite{Cheng:2000fu}, and are smaller by a factor of two in
the case of the orbifold considered here.

Consider the case $d = 6$. The electroweak observables are
logarithmically divergent at the one loop level, indicating that
within the framework of the effective $d$-dimensional electroweak
theory, they are unknown parameters to be fit to experiment. This
is reinforced by the higher loop estimates which are all of this
order if the cutoff is taken to be as large as possible -- where
the effective $d$-dimensional theory becomes strongly coupled. In
this case, the electroweak observables are directly sensitive to
the new physics at scales $M_s$ and above. It is possible, on the
other hand, that the cutoff is smaller or that the higher order
estimates are such that the one loop, logarithmic terms dominate.
Then the computations (\ref{T-parameter}), (\ref{sumS}), etc.,
enhanced relative to the 5-dimensional case by a large logarithm,
can be used to put a rough lower bound on $1/R$.

\begin{figure}[t]
\centerline{\epsfysize=7.5cm\epsfbox{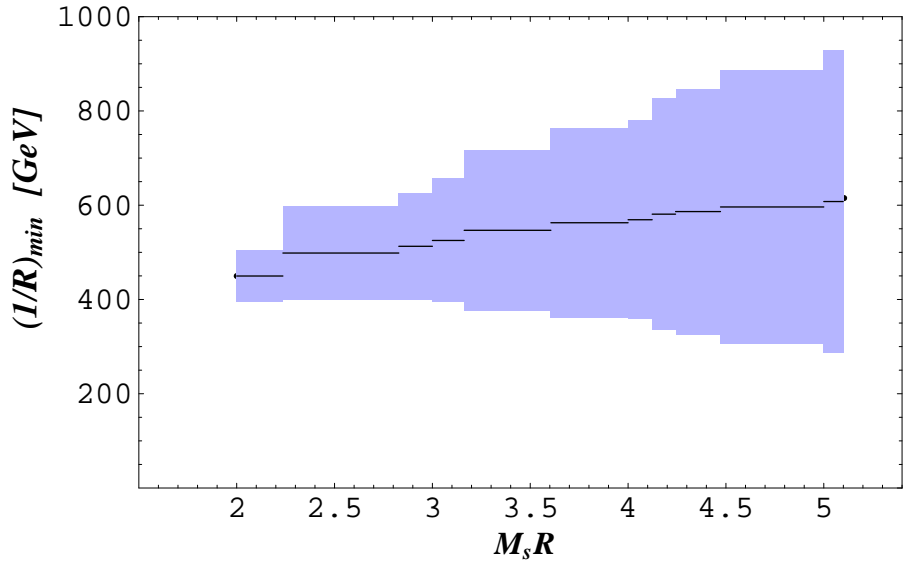}} \begin{center}
\parbox{5.5in}{ \caption[] {\small The lower bound on the
compactification scale as a function of the cut-off, for $\delta
=2$ extra dimensions. The vertical size of the shaded area is
given by the loop expansion parameter, $N_c \alpha_3(M_s) N_{\rm
KK}(M_s) / (4\pi)$, times the one-loop bound, and is a measure of
the theoretical uncertainty. For $M_sR \gae 5$ the standard model
interactions become non-perturbative, impeding a reliable estimate
of the electroweak observables. \label{6d}}} \end{center}
\end{figure}

In Fig.~1 we show the dependence of this lower bound on the ratio
between the cut-off $M_s$ and the compactification scale. Assuming
that the theory above $M_s$ is custodially symmetric, the one-loop
contribution to the $T$ parameter is reliable as long as the
theory remains perturbative, roughly for $M_sR \lae 5$. The KK
contributions to the one-loop coefficients of the $U(1)_Y$,
$SU(2)_W$ and $SU(3)_C$ $\beta$ functions are now $(81/10, \, 11/6
, \, -2)$, but again the $d$-dimensional $SU(3)_C$ interaction
becomes non-perturbative in the ultra-violet before the other
gauge interactions. The theoretical uncertainty due to higher
loops may be estimated in terms of the $\epsilon_3$ loop expansion
parameter. Fig.~1 shows that the lower bound on $1/R$ is increased
by roughly a factor of two compared to the 5-dimensional case, to
approximately $400 - 800$ GeV.

For $d \geq 7$, the cutoff dependence is more severe. The one-loop
estimate (3.8) for the $T$ parameter, for example, is enhanced by
the factor $(R M_s )^{d-6}$ relative to the 5-dimensional
estimate. Higher loop estimates are of the same order if the
ultraviolet cutoff, $M_s$, is above the scale where the couplings
become nonperturbative. Clearly, no reliable estimate is possible
in this case. For smaller $M_s$, the one-loop result has a strong
dependence on the cutoff, but otherwise the corrections are
smaller because the higher-dimensional operators have coefficients
suppressed by $1/(M_sR)^\delta$.

\section{Prospects for discovering Kaluza-Klein modes}
\setcounter{equation}{0}

We have shown that in the case of one universal extra dimension,
accessible to all the standard model fields, the fit to the
electroweak data allows KK excitations as light as 300 GeV. Such a
low bound raises the tantalizing possibility of discovering KK
states in the upcoming collider experiments.

If the KK number conservation is exact (that is, there is no
additional interaction violating the momentum conservation in
extra dimensions,) some of the KK excitations of the standard
model particles will be stable. The heavy-generation fermion KK
modes can decay to light generation fermion KK modes, {\it e.g.},
$b^{(1)} \to s^{(1)} (d^{(1)})\, + \, \gamma$, and the $W, Z$
gauge boson KK modes can decay to lepton KK modes and neutrinos,
$W^{(1)}(Z^{(1)}) \to \ell^{(1)} (\nu^{(1)})\, +\, \nu$. The KK
modes of the photon, gluon, and the lightest generation fermions
are stable and degenerate in mass, level-by-level, to a very good
approximation. Heavy stable charged particles will cause
cosmological problems if a significant number of them survive at
the time of nucleosynthesis~\cite{DeRujula:1990fe}.
For example, they will combine with other nuclei to form heavy
hydrogen atoms. Searches for such heavy isotopes put strong limits
on their abundance. Various cosmological arguments exclude these
particles with masses in the range of 100 GeV to 10 TeV, unless
there is a low scale inflation that dilutes their
abundance~\cite{Randall:1995fr,Lyth:1996ka}. The cosmological
problems can be avoided if there exist some KK-number-violating
interactions so that the non-zero KK states can decay. The
lifetime depends on the strengths of these KK number violating
interactions, which are usually suppressed by the cutoff scale
and/or the volume factor of the extra dimensions. For collider
searches, there is no difference between a stable particle and a
long-lived particle which decays outside the detector. We first
assume that the KK states are stable or long-lived. The case in
which the KK states decay promptly will then be considered when we
discuss the possible KK number violating interactions. \medskip

\subsection{Stable or Long-Lived KK Modes}

Because of the KK number conservation, the KK states have to be
produced in pairs or higher numbers.  They can only be produced at
LEP if their masses are less than $E_{\rm CM}/2,\, \sim 100$ GeV.
The current lower bound on the size of the extra dimensions is set
by the CDF and D0 experiments based on the Run I of the Tevatron.
The largest production cross-section is that for KK quarks and
gluons. After being produced, they will hadronize into
integer-charged states. Because of the large mass, they will be
slowly moving and the signatures are highly ionizing tracks.

For one extra dimension of radius $R$, the number of the quark KK
modes at each level is twice that of zero-modes, so neglecting the
light quark masses, there are six KK quarks of electric charge
$-1/3$ and mass $1/R$, four KK quarks of electric charge $2/3$ and
mass $1/R$, and two KK top-quarks of mass $\sqrt{1/R^2 + m_t^2}$.
Therefore, the production cross section for a pair of charged
tracks is roughly ten times higher than the one for a $q\bar{q}$
pair of quarks of mass $1/R$, $\sigma_{q\bar{q}}(1/R)$. For $1/R =
300$ GeV, $\sigma_{q\bar{q}}(1/R)\approx 0.1$ pb
\cite{Ellis:1991jf}.

The current lower mass limits on heavy stable quarks are 195 GeV
for charge 1/3 and 220 GeV for charge 2/3 \cite{Connolly:1999dv}.
The reach in mass would be approximately the same for two
charge-1/3 quarks as for one charge-2/3 quark. Hence, the current
bound on $1/R$ may be approximated as the mass limit on a
charge-2/3 quark, but with a production cross section about seven
times larger\footnote{The gluon KK modes further increase this
effective cross-section. The production cross-section for a pair
of gluon KK modes is larger than for a pair of quark KK modes, but
the probability for hadronizing into a charged meson is
significantly for a gluon KK mode.}. A dedicated study, beyond the
scope of this paper, is required to find this bound precisely.
However, by naively extrapolating the mass reach given in
\cite{Connolly:1999dv}, we estimate the lower direct bound on
$1/R$ to be in the $300-350$ GeV range.

It is remarkable that the direct lower bound on $1/R$ competes
with or even exceeds the indirect bound set by the electroweak
precision measurements. This should be contrasted with the case of
non-universal extra dimensions, where the non-conservation of the
KK number makes the indirect bound on $1/R$ stronger than the
direct one by a factor of five or so. Thus, Run II at the Tevatron
will either discover an universal extra dimension or else it will
significantly increase the lower bound on compactification scale.

\subsection{Short-Lived KK Modes}

 As mentioned above, the KK
states can decay into ordinary standard model particles if KK
number violating effects are present. Such violations of the KK
number can occur naturally. For example, the space in which the
standard model fields propagate may be a thick brane embedded in a
larger space in which gravitons propagate. In this case, the
standard model KK excitations can decay into standard model
particles plus gravitons going out of the thick brane (or other
neutral fields that can propagate outside the thick brane). The
unbalanced momentum in extra dimensions can be absorbed by the
thick brane. The lifetime depends on the strength of the coupling
to the particle going out of the brane and the density of its KK
modes (which depends on the volume of the space outside the thick
brane). If the KK states produced at the colliders decay promptly
inside the detector, the signatures will involve missing energy
and will be similar to supersymmetry. We assume that the KK number
violating interactions are {\it not} large enough to induce a
significant single-KK-state production cross section.

For the KK quark and gluon searches, the signature is multi-jets
plus missing energy, similar to the squarks and gluinos. At the
Tevatron Run I, the lower limits of the squark and gluino mass for
the equal mass case are 225 GeV at CDF~\cite{Done,Abel:2000vs} and
260 GeV at D0~\cite{Abbott:1999xc,Abel:2000vs}. The production
cross-sections of the KK quarks and gluons are similar to those of
the squarks and gluinos. The distributions of the jet energies and
the missing transverse energy however will depend on the masses of
the KK gravitons, {\it i.e.}, the size of the space outside the
thick brane.
We expect that the reaches in KK quarks and gluons are comparable
to those for squarks and gluinos in supersymmetric models. Run II
of the Tevatron is expected to probe squark and gluino masses up
to 350--400 GeV~\cite{Abel:2000vs}, so it could also probe KK
quarks and gluons beyond the current indirect limit in this
scenario. To distinguish the KK states from supersymmetry,
however, would require more detailed studies.

Another possibility for KK number violation is that there exist
some localized interactions of the standard model fields at a 3+1
dimensional subspace (3-brane) on the boundary or parallel to the
boundary. In the effective $d$-dimensional theory, these would
take the form of higher dimensional operators suppressed by powers
of the cutoff $M_s$. Some examples are \bear \label{localint} &&
\int dx^4 dy \delta (y-y_0) \frac{\lambda}{M_s} \overline{\Psi}
\not{\!\!D} \Psi ~, \nonumber \\ [2mm] && \int dx^4 dy \delta
(y-y_0) \frac{\lambda'}{M_s^{5/2}} \overline{\Psi} \sigma_{\alpha
\beta} F^{\alpha \beta} \Psi ~, \nonumber \\ [2mm] && \int dx^4 dy
\delta (y-y_0) \frac{\lambda''}{M_s^4} (\overline{\Psi}\Gamma_A
\Psi)(\overline{\Psi}\Gamma^A \Psi), \label{ops} \eear where
$\Psi$, $F^{\alpha \beta}$ are five-dimensional standard model
fermion and gauge fields, and $\Gamma_A$ is some combination of
the $\gamma$ matrices. The first contributes to the kinetic terms
of the KK states, so the KK mass spectrum would be modified after
re-diagonalizing and rescaling the kinetic terms into the
canonical form~\cite{Barbieri:2000vh}. The corrections (\ref{ops})
are suppressed by $M_s$ and we assume that the coefficients
($\lambda,\, \lambda^\prime, \, \lambda^{\prime\prime}, ...$) are
small enough so that these operators do not affect our analysis of
electroweak observables. However, they could be sufficiently large
to allow decays within the detector of the pair-produced KK modes.
The decay channels depend on which KK number violating
interactions are present. We discuss the simplest two-body decays
which can be induced by, {\it e.g.}, the first two interactions in
eq.~(\ref{localint}).

If the interactions involving the gluon field dominate, the KK
quarks and gluons decay into jets. The signals would be multi-jets
which are difficult to extract from the QCD backgrounds at the
Tevatron. However, if the interactions involving the electroweak
gauge bosons are large enough so that the decay of the KK quarks
into electroweak gauge bosons and quark zero-modes has a
significant branching ratio, we can invoke the searches for the
heavy quarks. For the decay into the $W$ bosons, the signal is
similar to the top quark. One can use the measurements of the top
quark production cross section at the
Tevatron~\cite{Abachi:1997re,Abe:1998iz} to put limits on the new
heavy quarks. In Ref.~\cite{Popovic:2000dx}, Popovic and Simmons
derived the bounds $\sigma^{q_H} (B_W)^2 < 7.8\, {\rm pb}\,
(12.0\, {\rm pb})$ at D0 (CDF), where $\sigma^{q_H}$ is the
cross-section for the heavy quark production, and $B_W$ is the
branching ratio for the heavy quark decaying to the $W$ boson and
an ordinary quark. Applying this result, we have $1/R \gae 200$
GeV for $B_W \sim 50\%$. There is also a search for the fourth
generation $b^\prime$ quark through the decay mode $ZZb\bar{b}$ at
CDF, which excluded the $b^\prime$ quark mass between 100 and 199
GeV if the branching ratio is 100\%~\cite{Affolder:2000bs}. This
can also apply to the KK states of the $b$ quark. In Run II at the
Tevatron, the decays of quark KK modes into a quark zero-mode and
a photon may be also significant. Other processes, potentially
relevant for Run II, include the electroweak production of a pair
of lepton KK modes with each of them subsequently  decaying into a
lepton zero-mode and a photon or a $W^\pm$, and the production of
a pair of KK modes of the electroweak gauge bosons leading to a
four-lepton signal. In general, the direct bounds on $1/R$ are
weaker and model dependent in this case.

With more extra dimensions the production cross section is higher
because of the multiplicity of KK modes. For example, with two
extra dimensions there are twice as many KK modes of mass $1/R$
than in the case of one extra dimension. However, the indirect
bounds may also be significantly higher. It is not clear whether
they are within the reach of Run II. The sensitivity of the LHC,
though, should be impressive, above a few TeV.

\section{Summary} \setcounter{equation}{0}

We have examined the experimental consequences of higher
dimensional theories in which all the standard model fields
propagate in the extra dimensions. With these ``universal'' extra
dimensions, contributions to precision electroweak observables
arise first at the one-loop level. In the case of a single extra
dimension, where the one-loop computations can be done reliably
within the framework of the effective 5-dimensional standard
model, the electroweak observables were estimated to allow a
compactification scale as low as $300$ GeV. We then noted that the
current lower bound from direct production experiments is set by
CDF and D0 to be in the few-hundred GeV range. Thus Run II at the
Tevatron will either see evidence for the extra dimensions or
significantly raise the lower bound on the compactification scale.

In the case of two universal extra dimensions, the electroweak
observables become logarithmically sensitive, at one loop, to the
cutoff on the effective 6-dimensional theory. If the cutoff is
taken to be as large as
 possible, where this effective theory becomes strongly coupled, then
the theory cannot be used to compute reliably the electroweak
observables. If, on the other hand, the cutoff is lower, with no
important
 contributions to the electroweak observables from higher scales, then
the observables may be estimated reliably at the one-loop level.
As indicated in Fig. 1, with $M_{s} R  \leq  5$, the lower bound
on the compactification scale is estimated to be between $400$ GeV
and $800$ GeV.

Besides opening the possibility of experimental detection of
universal extra dimensions in the near future, the lower bound on
the compactification scale discussed here suggests that physics in
extra dimensions may be responsible for electroweak symmetry
breaking. For example, standard model gauge interactions may
produce a bound-state Higgs doublet in six dimensions
\cite{Arkani-Hamed:2000hv} without excessive fine-tuning.

\bigskip

 {\bf Acknowledgements:} \ We would like to thank Mike Albrow,
 Zacharia Chacko, Lance Dixon, Jonathan Feng, Sheldon Glashow,
 Kevin Lynch, Eduardo Ponton, Marko Popovic, Erich Poppitz,
 Martin Schmaltz, Elizabeth Simmons, Matt Strassler, and Neal Weiner
 for helpful conversations and comments.
 The work of T. Appelquist and B. A. Dobrescu was supported by DOE under
contract DE-FG02-92ER-40704. H.-C. Cheng is supported by the
Robert R. McCormick Fellowship and by DOE Grant
DE-FG02-90ER-40560.

 \vfil 
\begin{thebibliography}{99} \frenchspacing

\bibitem{dudas} K.R.~Dienes, E.~Dudas and T.~Gherghetta, ``Extra
spacetime dimensions and unification,'' Phys.\ Lett.\ {\bf B436},
55 (1998) hep-ph/9803466.

\bibitem{Antoniadis:1990ew} I.~Antoniadis, ``A Possible New
Dimension At A Few TeV,'' Phys.\ Lett.\  {\bf B246}, 377 (1990).

\bibitem{Arkani-Hamed:2000dc} N.~Arkani-Hamed and M.~Schmaltz,
``Hierarchies without symmetries from extra dimensions,'' Phys.\
Rev.\ {\bf D61}, 033005 (2000) hep-ph/9903417.

\bibitem{Arkani-Hamed:2000hv} N.~Arkani-Hamed, H.-C.~Cheng,
B.~A.~Dobrescu and L.~J.~Hall, ``Self-breaking of the standard
model gauge symmetry,'' Phys.\ Rev.\  {\bf D62}, 096006 (2000)
hep-ph/0006238.

\bibitem{bounds} W.~J.~Marciano, ``Precision electroweak
measurements and 'new physics','' hep-ph/9902332 and ``Fermi
constants and 'new physics','' Phys.\ Rev.\  {\bf D60}, 093006
(1999), hep-ph/9903451; \\ P.~Nath and M.~Yamaguchi, ``Effects of
extra space-time dimensions on the Fermi constant,'' Phys.\ Rev.\
{\bf D60}, 116004 (1999), hep-ph/9902323; \\ M.~Masip and
A.~Pomarol, ``Effects of SM Kaluza-Klein excitations on
electroweak observables,'' Phys.\ Rev.\  {\bf D60}, 096005 (1999),
hep-ph/9902467; \\ T.~G.~Rizzo and J.~D.~Wells, ``Electroweak
precision measurements and collider probes of the standard  model
with large extra dimensions,'' Phys.\ Rev.\  {\bf D61}, 016007
(2000), hep-ph/9906234; \\  A.~Strumia, ``Bounds on Kaluza-Klein
excitations of the SM vector bosons from electroweak tests,''
Phys.\ Lett.\  {\bf B466}, 107 (1999), hep-ph/9906266; \\
R.~Casalbuoni, S.~De Curtis, D.~Dominici and R.~Gatto, ``SM
Kaluza-Klein excitations and electroweak precision tests,'' Phys.\
Lett.\  {\bf B462}, 48 (1999), hep-ph/9907355; \\ C.~D.~Carone,
``Electroweak constraints on extended models with extra
dimensions,'' Phys.\ Rev.\  {\bf D61}, 015008 (2000),
hep-ph/9907362; \\ A.~Delgado, A.~Pomarol and M.~Quiros,
``Electroweak and flavor physics in extensions of the standard
model with  large extra dimensions,'' JHEP {\bf 0001}, 030 (2000),
hep-ph/9911252.



\bibitem{Green:1984sg} M.~B.~Green and J.~H.~Schwarz, ``Anomaly
Cancellation In Supersymmetric D=10 Gauge Theory And Superstring
Theory,'' Phys.\ Lett.\  {\bf B149}, 117 (1984).

\bibitem{Cheng:2000fu} H.-C.~Cheng, B.~A.~Dobrescu and C.~T.~Hill,
``Gauge coupling unification with extra dimensions and
gravitational  scale effects,'' Nucl.\ Phys.\  {\bf B573}, 597
(2000), hep-ph/9906327.

\bibitem{Appelquist:1993ka} T.~Appelquist and G.~Wu, ``The
Electroweak chiral Lagrangian and new precision measurements,''
Phys.\ Rev.\  {\bf D48}, 3235 (1993), hep-ph/9304240.


\bibitem{Chivukula:2000px} R.~S.~Chivukula, C.~Holbling and
N.~Evans, ``Limits on a composite Higgs boson,'' Phys.\ Rev.\
Lett.\  {\bf 85}, 511 (2000), hep-ph/0002022.

\bibitem{ewwg}The LEP Electroweak Working Group, \\
http://lepewwg.web.cern.ch/LEPEWWG/plots/summer2000/

\bibitem{DeRujula:1990fe} A.~De Rujula, S.~L.~Glashow and
U.~Sarid, ``Charged Dark Matter,'' Nucl.\ Phys.\  {\bf B333}, 173
(1990); \\
S.~Dimopoulos, D.~Eichler, R.~Esmailzadeh and G.~D.~Starkman,
``Getting A Charge Out Of Dark Matter,'' Phys.\ Rev.\  {\bf D41},
2388 (1990); \\
R.~S.~Chivukula, A.~G.~Cohen, S.~Dimopoulos and T.~P.~Walker,
``Bounds On Halo Particle Interactions From Interstellar
Calorimetry,'' Phys.\ Rev.\ Lett.\  {\bf 65}, 957 (1990); \\
A.~Gould, B.~T.~Draine, R.~W.~Romani and
S.~Nussinov, ``Neutron Stars: Graveyard Of Charged Dark Matter,''
Phys.\ Lett.\  {\bf B238}, 337 (1990).


\bibitem{Randall:1995fr} L.~Randall and S.~Thomas, ``Solving the
cosmological moduli problem with weak scale inflation,'' Nucl.\
Phys.\  {\bf B449}, 229 (1995), hep-ph/9407248.



\bibitem{Lyth:1996ka} D.~H.~Lyth and E.~D.~Stewart, ``Thermal
inflation and the moduli problem,'' Phys.\ Rev.\  {\bf D53}, 1784
(1996), hep-ph/9510204.




\bibitem{Ellis:1991jf} R.~K.~Ellis, ``Rates for top quark
production,'' Phys.\ Lett.\  {\bf B259}, 492 (1991).




\bibitem{Connolly:1999dv} A.~Connolly  [CDF collaboration],
``Search for long-lived charged massive particles at CDF,'' Talk
at the American Physical Society (APS) Meeting of the Division of
Particles and Fields (DPF 99), Los Angeles, CA, Jan 5-9, 1999,
hep-ex/9904010.


\bibitem{Done} J.~P.~Done [CDF collaboration], Talk at the
American Physical Society Centennial Meeting, Atlanta, GA, March
20--26, 1999.


\bibitem{Abel:2000vs} S.~Abel {\it et al.}  [SUGRA Working Group
Collaboration], ``Report of the SUGRA working group for run II of
the Tevatron,'' hep-ph/0003154.



\bibitem{Abbott:1999xc} B.~Abbott {\it et al.}  [D0
Collaboration], ``Search for squarks and gluinos in events
containing jets
 and a large  imbalance in transverse energy,''
Phys.\ Rev.\ Lett.\  {\bf 83}, 4937 (1999), hep-ex/9902013.

\bibitem{Barbieri:2000vh} R.~Barbieri, L.~J.~Hall and Y.~Nomura,
``A Constrained Standard Model from a Compact Extra Dimension,''
hep-ph/0011311.


\bibitem{Abachi:1997re} S.~Abachi {\it et al.}  [D0
Collaboration], ``Measurement of the top quark pair production
cross section in p anti-p  collisions,'' Phys.\ Rev.\ Lett.\  {\bf
79}, 1203 (1997), hep-ex/9704015.



\bibitem{Abe:1998iz} F.~Abe {\it et al.}  [CDF Collaboration],
``Measurement of the top quark mass and t anti-t production cross
section  from dilepton events at the Collider Detector at
Fermilab,'' Phys.\ Rev.\ Lett.\  {\bf 80}, 2779 (1998),
hep-ex/9802017.



\bibitem{Popovic:2000dx} M.~B.~Popovic and E.~H.~Simmons,
``Weak-singlet fermions: Models and constraints,'' Phys.\ Rev.\
{\bf D62}, 035002 (2000), hep-ph/0001302.



\bibitem{Affolder:2000bs} T.~Affolder {\it et al.}  [CDF
Collaboration], ``Search for a fourth-generation quark more
massive than the Z0 boson in  p anti-p collisions at $\sqrt{s}$ =
1.8 TeV,'' Phys.\ Rev.\ Lett.\ {\bf 84}, 835 (2000),
hep-ex/9909027.

 \end{thebibliography}
\end{document}